\documentstyle[12pt,aaspp4]{article}
\begin{document}

\title{A Correlation Between Inclination and Color in the Classical
Kuiper Belt}
\author{Chadwick A. Trujillo}
\affil{California Institute of Technology, Division of Geological and
Planetary Sciences, MS 150-21, Pasadena, CA  91125 \\
chad@gps.caltech.edu}
\and
\author{Michael E. Brown}
\affil{California Institute of Technology, Division of Geological and
Planetary Sciences, MS 150-21, Pasadena, CA  91125 \\
mbrown@caltech.edu}

\begin{abstract}
We have measured broadband optical $BVR$ photometry of 24 Classical
and Scattered Kuiper belt objects (KBOs), approximately doubling the
published sample of colors for these classes of objects.  We find a
statistically significant correlation between object color and
inclination in the Classical Kuiper belt using our data.  The color
and inclination correlation increases in significance after the
inclusion of additional data points culled from all published works.
Apparently, this color and inclination correlation has not been more
widely reported because the Plutinos show no such correlation, and
thus have been a major contaminant in previous samples.  The color and
inclination correlation excludes simple origins of color diversity,
such as the presence of a coloring agent without regard to dynamical
effects.  Unfortunately, our current knowledge of the Kuiper belt
precludes us from understanding whether the color and inclination trend
is due to environmental factors, such as collisional resurfacing, or
primordial population effects.  A perihelion and color correlation is
also evident, although this appears to be a spurious correlation
induced by sampling bias, as perihelion and inclination are correlated
in the observed sample of KBOs.
\end{abstract}

Subject headings: comets: general --- Kuiper Belt, Oort Cloud --- solar
system: formation

\section{Introduction}

Beyond Neptune, the Solar System is populated by $\sim 10^{5}$ Kuiper
belt objects (KBOs, also trans-Neptunians) larger than 100 km in
diameter, which represent about 0.1 Earth masses of material and are
presumed to contain the least thermally processed material in the
solar system (Trujillo, Jewitt \& Luu 2001).  As of Oct 2001, $\approx
480$ KBOs are known, less than 1\% of the inferred population.
Dynamically, the KBOs can be divided into at least three classes: (1)
the Classical KBOs represent 2/3 of the observed sample and about 1/2
the inferred population; (2) the Resonant KBOs are comprised primarily
of Plutinos ($a \approx 39.4$ AU) and objects in 2:1 mean-motion
resonance with Neptune ($a \approx 47.7$ AU), representing 15\% of the
observed sample, but only 6\% of the total population; and (3) the
Scattered KBOs comprise about 10\% of the observed sample, but nearly
half the inferred population because their extreme dynamical
properties ($50 \mbox{ AU} < a < 200 \mbox { AU}$ and eccentricities
$0.2 < e < 0.9$) place them at unobservably great heliocentric
distances for large fractions of their orbits (Trujillo, Jewitt \& Luu
2000).

Since the KBOs are thought to be the least thermally modified bodies
in the solar system, they are expected to provide insight into the
physical and chemical processes that dominated the primitive solar
nebula.  Given the fact that the KBOs have been observed for nearly a
decade, very little is known about their physical properties,
primarily because of their extreme faintness (median $R$ magnitude
$m_{R} = 23.0$).  The most striking physical feature observed of the
KBOs is their substantial color diversity, from nearly neutral
reflectance to the reddest objects in the solar system (Jewitt \& Luu
1998 and Tegler \& Romanishin 2000).  Objects are reddened from the
visible throughout the near-IR (Davies et al. 2000), presumably from a
single coloring agent (Jewitt and Luu 2001), although the physical and
chemical mechanisms for the reddening process have not been
identified.  Recently, Tegler \& Romanishin (2000) reported that the
Classical KBOs exhibited a correlation between color and inclination.
No other observers have reported this trend, and Tegler \& Romanishin
estimate the significance of their reported trend using their
two-color paradigm, which has not been confirmed by independent
sources (Barucci et al. 2001 and Jewitt \& Luu 2001).  Thus, we chose
to conduct our own $BVR$ observations of this population.  Using our
new observations, we demonstrate that the inclination and color
correlation is indeed statistically significant.  In addition, we note
the presence of a perihelion and color correlation which appears to be
due to sampling bias.

\section{Observations}
\label{observations}

We chose to observe the Classical KBOs --- those KBOs which are
expected to be the least dynamically modified by Neptune.  In
practice, 19 Classical KBOs and 5 Scattered KBOs ($a > 50$ AU, 30 AU $
< q < $ 40 AU) were observed due to the distribution of objects in the
sky when observations were scheduled.  Results of this work are not
changed significantly by the inclusion or exclusion of the Scattered
KBOs, as they are a minority population in our sample.  All data were
taken with the same equipment, the LRIS (Oke et al. 1995) and LRIS-B
(McCarthy et al. 1998) cameras at the 10 meter Keck I telescope during
the nights of Feb 26--27, 2001.  This dichroic-based instrument
combination allows the simultaneous use of blue and visible filters
(yielding $B-V$ color) or blue and red filters (yielding $B-R$ color),
allowing a factor two increase in efficiency over the previous single
camera system.  The efficiency of the LRIS/LRIS-B instrument has
permitted us to approximately double the published sample of broadband
optical colors for the Classical KBOs in just two nights.

\begin{figure}
\plotfiddle{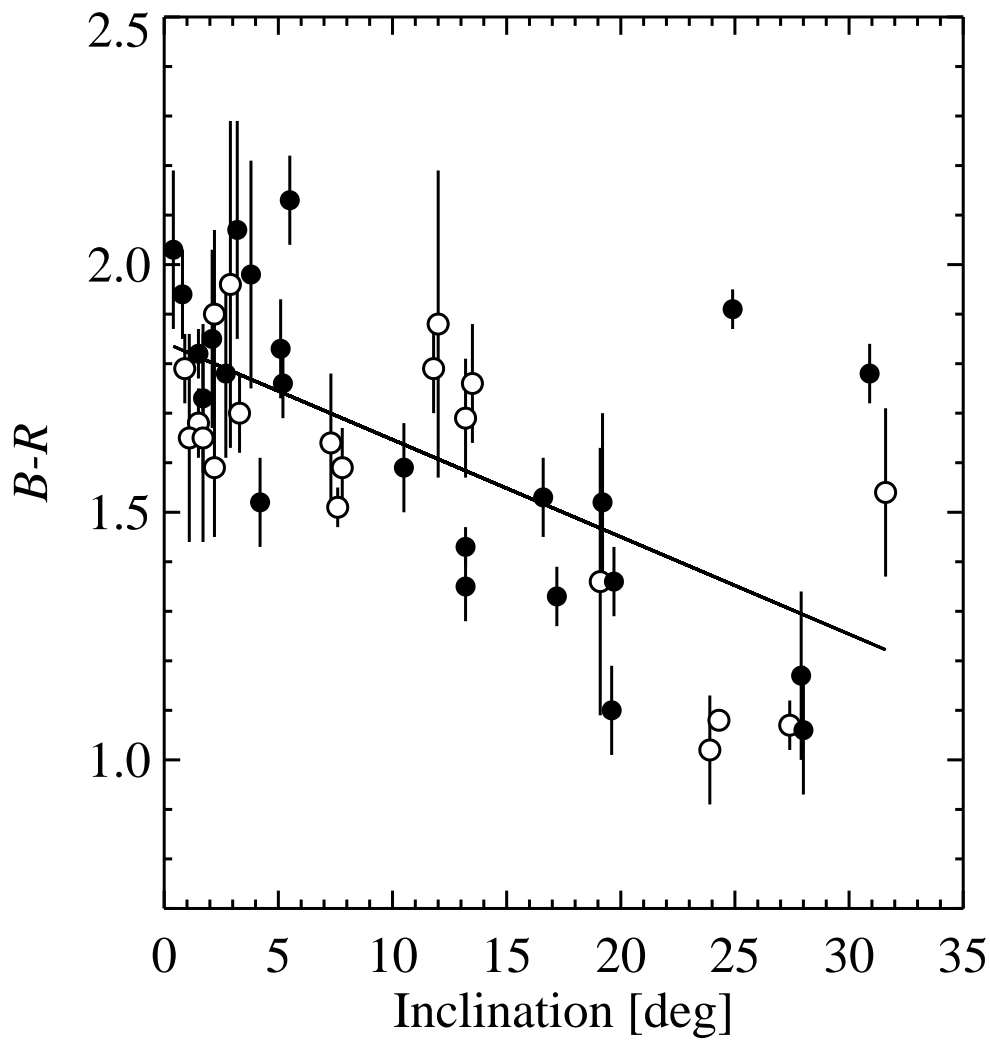}{6in}{0}{80}{80}{-250}{-100}
\figcaption{$B-R$ color vs. inclination for the Classical KBOs and
Scattered KBOs in our sample (filled circles) and all published data
(no Plutinos, hollow circles).  A linear least-squares fit has also
been plotted for illustrative purposes (line, slope=$-0.0196 \pm
0.003$ magnitudes/deg, intercept=$1.84 \pm 0.5$ magnitudes).  The
trend corresponds to $3.1 \sigma$ (our data points) and $4.1 \sigma$
(all data points) significance, as estimated by the Spearman rank
correlation method.
\label{inclbmr}}
\end{figure}

Standard stars (Landolt 1992) were imaged at a variety of airmasses
and photometric reductions were performed using a two dimensional
PSF-fitting procedure, which allows for arbitrary non-radially
symmetric variations of the PSF.  Using this method, $> 20$ bright
field stars were sampled in both spatial dimensions of the focal plane
to estimate the true point-spread-function (PSF) for a given image.
Since exposure times were fixed at 5 minutes, KBO apparent motion (3
arc-sec/hr or 0.25 arc-sec/exposure) was minimized, yielding no
significant difference between the KBO PSFs and the stellar PSFs given
the $\sim 1$ arc-sec $B$-band seeing (seeing for the $R$ and $V$ bands
was similar, since focus was optimized for $B$).  The stellar PSF was
fit to the KBO image to determine flux and local background levels.
In all cases, KBO PSFs were consistent with the model PSFs within the
Poisson noise.  Additionally, synthetic data tests were performed to
demonstrate that the PSF fitting procedure is robust to errors in KBO
centroiding.  Approximately 4 $B$, 2 $V$ and 2 $R$ measurements were
taken of each KBO, although measurements were repeated if
contamination by background sources was suspected.  The reduced data
appear free of instrumental biases --- no correlations were observed
between color and data collection parameters such as airmass, seeing,
time of observations, apparent motion of objects, location in the
focal plane or sun-target-observer angle.  The most striking feature
of these data appears in Figure~\ref{inclbmr}, which shows an apparent
correlation between orbital inclination $i$ and $B-R$ color.  With few
exceptions, this correlation appears to vary continuously from the
reddest (low-inclination) KBOs to the bluest (high-inclination) KBOs.
Both the $B-R$ and $B-V$ color measurements for each object are
presented in Table~\ref{ourdata}.  The physical implications of these
observations are discussed in \S~\ref{interpretation}, while an
estimation of the statistical significance of the color and
inclination trend, as well as other trends, appears in the remainder
of this section.

\subsection{The KBO Inclination and Color Correlation}

To estimate the significance of the relationship between color and
$i$, we computed the Spearman (1904) rank correlation statistic $r$
for our data set, which takes on non-zero values between $-1 < r < 1$
for correlated quantities.  The strength of the Spearman method is
that it does not presuppose a functional form for the correlation, as
do linear least-squares estimates.  In addition, the Spearman method
is less sensitive to the presence of deviant data points than a
chi-squared rank correlation test, as it is the difference in rank
between sorted data lists that is compared, not the actual data values
themselves.  Two statistical tests (Press et al. 1992), the
$t$-statistic and the variance of the sum squared difference of ranks,
$\mbox{Var}(D)$, can be used to estimate the significance of a
non-zero value of $r$ (here, we report the mean of the two tests).  We
measure the correlation between $B-R$ color (a stronger indicator of
color than $B-V$) and orbital inclination as $r = -0.62$ for the $N =
24$ data points in our new sample.  The probability of this $r$ or a
more significant one occurring in an uncorrelated sample is $P = 2.0
\times 10^{-3}$ ($3.1 \sigma$ significance, assuming Gaussian
statistics, since $1-P$ yields the confidence level).

In order to increase sample size, we repeated this analysis combining
our data with all previously published data (Table~\ref{otherdata}).
Color measurements of the same object from multiple investigators have
been combined into one value, with error bars reflecting the greater
of (1) the standard error of the color values or (2) the mean of the
reported errors when combined in quadrature.  We note that the process
that produces the colors appears stochastic in nature, with individual
KBOs departing from the color correlation based on their past
interactions with the coloring mechanism.  Error bars reflect color
measurement error, and not the expected departure of an individual KBO
from the overall color / inclination trend, which cannot be computed a
priori without more information about the physical nature of the
coloring process.  We note that in cases of duplication between our
objects and published objects, our color values are statistically
indistinguishable from other published values.  By increasing the
sample size using previously published data on all KBOs, the
significance of the inclination and color correlation increases to
$3.3 \sigma$ ($N=60$).

It appears that the correlation between color and inclination has not
been widely reported because of the influence of the Plutinos, for
which no correlation is apparent.  Removing the Plutinos from the
combined sample of data results in a large increase in the correlation
statistic to $4.1 \sigma$ ($N=44$).  However, we must note that the
removal of the Plutinos is still statistically tentative, as a
two-dimensional Kolmogorov-Smirnov test (Press et al. 1992) shows that
the null hypothesis that the Plutino and other KBO ($i$, $B-R$) values
are drawn from the same distribution can be rejected with only 93\%
($<2 \sigma$) confidence.  This observation in itself has some
potential implications for the origin of color diversity, as discussed
in the next section.  Since the correlation between inclination and
color for the non-Plutino KBOs (1) increases in significance with
increased sample size, (2) does not appear to be the product of any
observational biases, (3) is observed in both $B-V$ and the $B-R$, and
(4) has $< 4 \times 10^{-5}$ ($4.1 \sigma$) chance of being produced
in an uncorrelated population, we conclude that it is statistically
significant.

\subsection{Other Correlations}
\label{othercorr}

\begin{figure}
\plotfiddle{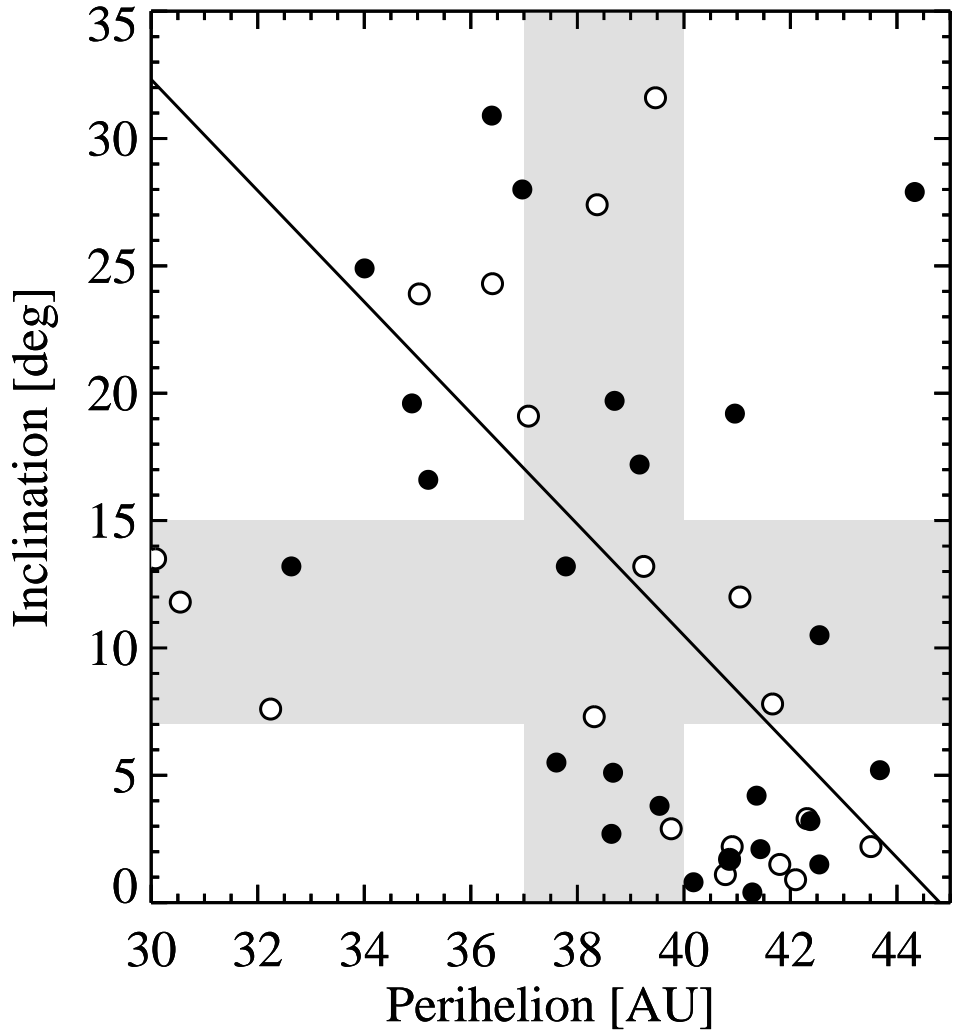}{6in}{0}{80}{80}{-250}{-100}
\figcaption{Inclination vs. perihelion for the Classical KBOs and
Scattered KBOs in our sample (filled circles) and all published data
(no Plutinos, hollow circles).  For illustrative purposes, a linear
least-squares fit has been plotted for points with perihelion $> 34$
AU (line).  The shaded bars represent the constant $q$ (vertical bar)
and constant $i$ (horizontal bar) subsamples detailed in Figure 3.
\label{periincl}}
\end{figure}

\begin{figure}
\plotfiddle{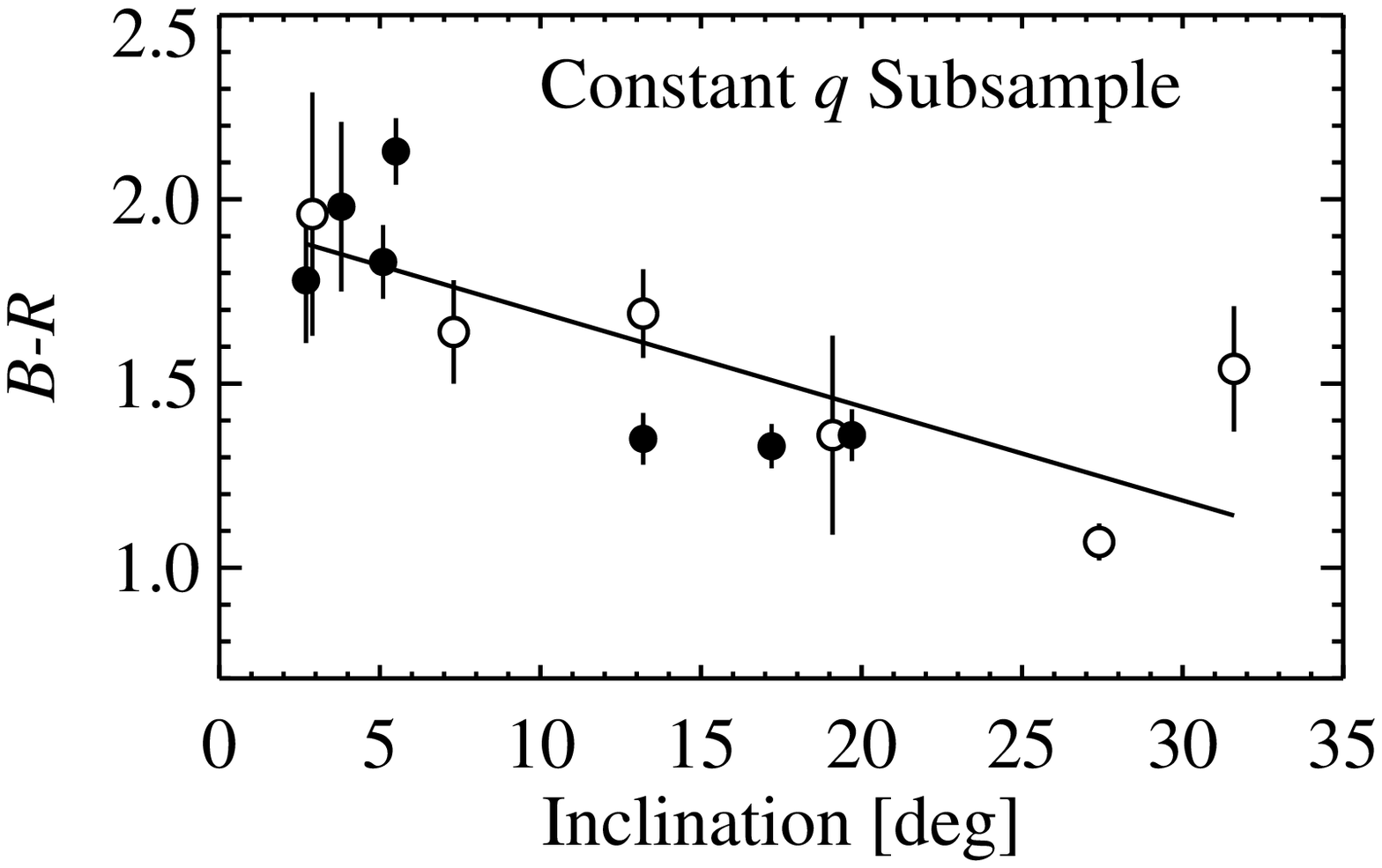}{3in}{0}{80}{80}{-250}{-100}
\plotfiddle{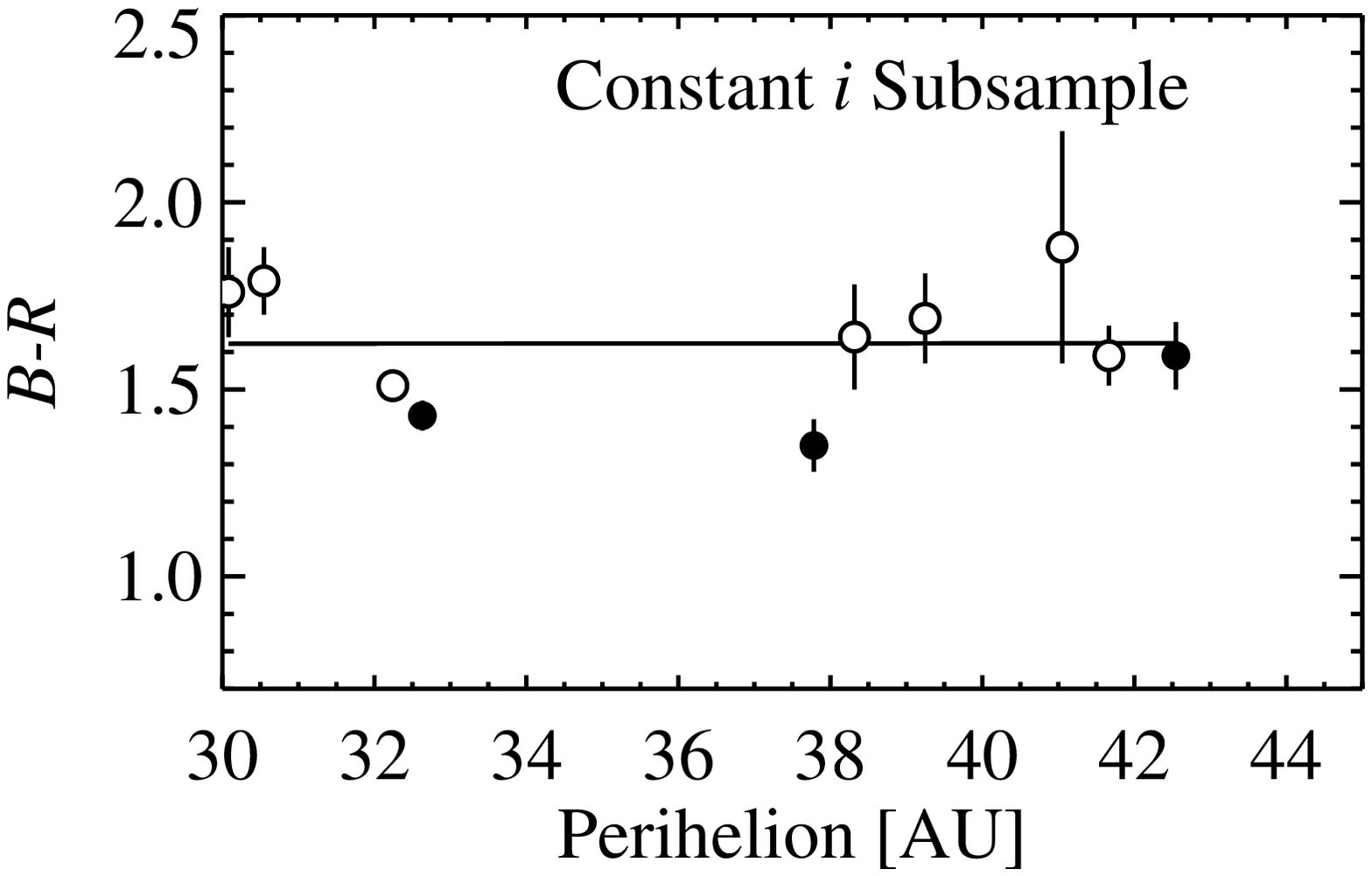}{3in}{0}{80}{80}{-250}{-120}
\figcaption{Color vs. inclination for the constant $q$ subsample (top)
and color vs. perihelion for the constant $i$ subsample (bottom).  The
top trend corresponds to $2.8 \sigma$, and the bottom trend is not
significant, as estimated by the Spearman rank correlation method.
The constant perihelion subsample (top) indicates that the
inclination/color correlation is not due to observational bias.  In
contrast, the constant $i$ subsample lacks any trend, indicating that
the observed perihelion/color correlation is most likely due to
sampling bias induced by the perihelion/inclination
correlation. \label{constant}}
\end{figure}

The only other color trend that we observe for the Classical KBOs is a
perihelion and color correlation.  This trend, however, is almost
certainly due to sampling bias.  To illustrate this, we have
considered two subsamples of the $B-R$ data, which are shown as the
greyed bars in Figure~\ref{periincl}.  The first subsample (the
``constant $q$'' subsample) shows little perihelion variation but
large inclination variation and consists of objects between $37\mbox{
AU} < q < 40\mbox{ AU}$.  The constant $q$ subsample exhibits a $2.8
\sigma$ correlation between inclination and color, indicating that the
inclination and color correlation cannot be attributed to perihelion
variations alone (Figure~\ref{constant}).  The second subsample (the
``constant $i$'' subsample) shows little inclination variation but
large perihelion variation and consists of objects between $7\arcdeg <
i < 15\arcdeg$.  The constant $i$ subsample shows no correlation
between perihelion and color, indicating that the perihelion and color
trend is most likely a spurious correlation induced by sampling bias.

\section{Interpretation}
\label{interpretation}

The link between color and inclination rules out very simple
mechanisms of producing the observed KBO color diversity.  The
presence of a coloring agent (surface frosts, for example) independent
of the Kuiper belt dynamical environment cannot be the only cause of
color diversity, because in such a scenario, no correlation between
color and any orbital parameter would be expected.  More complex
processes must be invoked such as evolutionary processes or remnant
primordial composition effects.

Collisional resurfacing has been suggested as a possible mechanism for
producing the wide range of KBO colors (Luu \& Jewitt 1996 and Jewitt
\& Luu 2001).  In this model, a close competition between radiative
processing, which reddens the surface, and collisional resurfacing,
which restores neutral ices to the surface, produces the observed
color variations.  Assuming that collisions are most likely to take
place near the ecliptic where the KBO number densities are highest, a
color and inclination correlation seems likely.  An object's speed
perpendicular to the ecliptic follows $\Delta v \propto \sin{i}$.
From laboratory simulations (Fujiwara, Kamimoto \& Tsukamoto 1977)
scaled to solar system events (Fujiwara et al. 1989), the mass ejected
during a collision is proportional to impact energy squared, $m
\propto E^{2} \propto \Delta v^{4}$.  Thus, ejecta mass could be a
strong function of object inclination, $m \propto \sin^{4}{i}$,
leading to a correlation between color and inclination.  If this were
the case, one might expect to see a correlation between color and
population averaged impact velocity $v_{\rm imp} \propto \sqrt{i^2 +
e^2}$ (Stern 1995).  Unfortunately, because of the paucity of color
data, it is not currently possible to separate the effects of an $i$
correlation from a $v_{\rm imp}$ correlation because the two
parameters are so strongly correlated.  In addition, it is not clear
why the collisional environment would be different for the Plutinos,
which show no correlations between $i$ and color.  Further modeling is
required to fully quantify this effect.  Unfortunately, the
competition between collisions and radiative processing is quite
difficult to model, as we do not have a thorough understanding of all
of the astrophysical effects that take place.  For instance, no
dynamical models have yet estimated collision rates based on
population dynamics in a manner that could demonstrate a
dynmaics-induced color correlation.  In addition, models of radiative
processing have not addressed the effect of the complete carbonization
of surfaces, which would result in a dark, neutrally colored surface.
Furthermore, no collisional resurfacing models have adequately
reproduced the magnitude of color variations presently observed for
the KBOs (Jewitt \& Luu 2001).

The best alternative to evolutionary processes is that the colors are
primordial in nature.  An $i$ correlation might be expected if the
KBOs were composed of multiple sub-populations with different
primordial colors and inclinations.  Such a model is compatible with
two recently reported dynamical groupings: (1) the Levison and Stern
(2001) large and small size components of the Classical KBOs and (2)
the Brown (2001) high-$i$ and low-$i$ components of the Classical
KBOs.  However, we find no correlations between color and any orbital
parameter other than $i$, as might be expected for a cosmogonic origin
to the colors.  Even if the KBO colors are dominated by primordial
processes, the role of collisions in altering KBO surfaces must still
be addressed given that (1) collisions are thought to be responsible
for reducing the mass of the Kuiper belt by a factor $\sim 100$ since
the epoch of formation (Stern \& Colwell 1997 and Kenyon \& Luu 1999)
and (2) it is expected that a large fraction of KBO surfaces are
covered with impact signatures (Durda \& Stern 2000).

\acknowledgments

Antonin Bouchez provided observational assistance at the telescope.
David Sprayberry and Gabrelle Saurage provided technical and
operational assistance at the telescope.

\newpage

\begin{center}
\begin{deluxetable}{lllrllll}
\small
\tablewidth{0pt}
\tablecaption{Photometric Data From This Work}
\tablehead{\colhead{Object} & \colhead{$a$\tablenotemark{a}} &
\colhead{$e$\tablenotemark{b}} & \colhead{$i$\tablenotemark{c}} &
\colhead{$q$\tablenotemark{d}} & \colhead{$R$} & \colhead{$B-R$} & \colhead{$B-V$} \\
\colhead{} & \colhead{[AU]} & \colhead{} & \colhead{[deg]} & \colhead{[AU]} &
\colhead{[magnitudes]} & \colhead{[magnitudes]}}
\startdata
1999 $\rm HU_{11}$                   & 43.955 & 0.061 &  0.4 &  41.285 & $23.04 \pm 0.06$ & $2.03 \pm 0.16$ & $1.36 \pm 0.16$ \nl
1999 $\rm CO_{153}$                  & 43.976 & 0.086 &  0.8 &  40.181 & $22.69 \pm 0.03$ & $1.94 \pm 0.09$ & $1.03 \pm 0.10$ \nl
1997 $\rm CU_{29}$\tablenotemark{e}  & 43.771 & 0.028 &  1.5 &  42.542 & $22.59 \pm 0.03$ & $1.82 \pm 0.05$ & $1.10 \pm 0.05$ \nl
1994 $\rm EV_{3}$\tablenotemark{e}   & 42.792 & 0.045 &  1.7 &  40.871 & $23.64 \pm 0.06$ & $1.73 \pm 0.15$ & $1.15 \pm 0.23$ \nl
2000 $\rm FS_{53}$                   & 42.960 & 0.035 &  2.1 &  41.437 & $23.29 \pm 0.05$ & $1.85 \pm 0.18$ & $1.23 \pm 0.20$ \nl
1999 $\rm CM_{119}$                  & 44.623 & 0.134 &  2.7 &  38.640 & $23.58 \pm 0.06$ & $1.78 \pm 0.17$ & $0.90 \pm 0.19$ \nl
1999 $\rm CJ_{119}$                  & 45.647 & 0.072 &  3.2 &  42.375 & $22.92 \pm 0.21$ & $2.07 \pm 0.22$ & $1.38 \pm 0.18$ \nl
2000 $\rm CM_{105}$                  & 42.406 & 0.068 &  3.8 &  39.543 & $22.46 \pm 0.03$ & $1.98 \pm 0.23$ & $1.14 \pm 0.23$ \nl
2000 $\rm CL_{105}$                  & 43.234 & 0.043 &  4.2 &  41.365 & $23.25 \pm 0.06$ & $1.52 \pm 0.09$ & $1.11 \pm 0.08$ \nl
1999 $\rm HT_{11}$                   & 43.602 & 0.113 &  5.1 &  38.670 & $23.05 \pm 0.04$ & $1.83 \pm 0.10$ & $1.14 \pm 0.14$ \nl
1999 $\rm GS_{46}$                   & 44.074 & 0.009 &  5.2 &  43.680 & $22.64 \pm 0.02$ & $1.76 \pm 0.07$ & $1.13 \pm 0.09$ \nl
1999 $\rm CV_{118}$                  & 53.152 & 0.292 &  5.5 &  37.607 & $22.78 \pm 0.06$ & $2.13 \pm 0.09$ & $1.08 \pm 0.08$ \nl
1999 $\rm JD_{132}$                  & 42.996 & 0.010 & 10.5 &  42.546 & $22.26 \pm 0.02$ & $1.59 \pm 0.09$ & $1.11 \pm 0.10$ \nl
1999 $\rm CQ_{133}$                  & 41.517 & 0.090 & 13.2 &  37.784 & $23.07 \pm 0.05$ & $1.35 \pm 0.07$ & $0.75 \pm 0.07$ \nl
1999 $\rm HB_{12}$                   & 55.359 & 0.411 & 13.2 &  32.631 & $22.12 \pm 0.02$ & $1.43 \pm 0.04$ & $0.88 \pm 0.06$ \nl
1999 $\rm CG_{119}$                  & 49.926 & 0.295 & 16.6 &  35.201 & $23.21 \pm 0.04$ & $1.53 \pm 0.08$ & $0.87 \pm 0.09$ \nl
1999 $\rm HW_{11}$                   & 52.422 & 0.253 & 17.2 &  39.168 & $22.93 \pm 0.03$ & $1.33 \pm 0.06$ & $0.80 \pm 0.08$ \nl
2000 $\rm CO_{105}$                  & 47.183 & 0.132 & 19.2 &  40.958 & $22.58 \pm 0.18$ & $1.52 \pm 0.18$ & $0.82 \pm 0.06$ \nl
\tablebreak
2000 $\rm CQ_{105}$                  & 57.105 & 0.389 & 19.6 &  34.895 & $23.02 \pm 0.05$ & $1.10 \pm 0.09$ & $0.71 \pm 0.09$ \nl
1999 $\rm CF_{119}$                  & 90.197 & 0.571 & 19.7 &  38.699 & $22.90 \pm 0.04$ & $1.36 \pm 0.07$ & $0.77 \pm 0.07$ \nl
1999 $\rm KR_{16}$\tablenotemark{e}  & 48.477 & 0.298 & 24.9 &  34.006 & $21.31 \pm 0.03$ & $1.91 \pm 0.04$ & $1.13 \pm 0.06$ \nl
2000 $\rm CG_{105}$                  & 46.595 & 0.049 & 27.9 &  44.334 & $23.40 \pm 0.16$ & $1.17 \pm 0.17$ & $0.78 \pm 0.09$ \nl
1998 $\rm HL_{151}$                  & 40.728 & 0.092 & 28.0 &  36.967 & $23.96 \pm 0.08$ & $1.06 \pm 0.13$ & $0.59 \pm 0.19$ \nl
2000 $\rm AF_{255}$                  & 49.350 & 0.263 & 30.9 &  36.394 & $22.92 \pm 0.03$ & $1.78 \pm 0.06$ & $1.16 \pm 0.07$ \nl
\enddata
\tablecomments{Objects are ordered by increasing inclination.  Orbital
elements were provided by the Minor Planet Center.}
\tablenotetext{a}{semimajor axis}
\tablenotetext{b}{eccentricity}
\tablenotetext{c}{$i$ refers to the inclination with respect to the
ecliptic.  Transforming the inclination to the invariable plane (mean
correction 0.6\arcdeg, Burkhardt 1982) does not have a significant
impact on results.}
\tablenotetext{d}{perihelion distance}
\tablenotetext{e}{These objects were also measured by another group,
see Table 2.}
\label{ourdata}
\end{deluxetable}
\end{center}

\newpage

\begin{center}
\begin{deluxetable}{lllrllll}
\small
\tablewidth{0pt}
\tablecaption{Photometric Data From All Other Published Works}
\tablehead{\colhead{Object} & \colhead{$a$\tablenotemark{a}} &
\colhead{$e$\tablenotemark{b}} & \colhead{$i$\tablenotemark{c}} &
\colhead{$q$\tablenotemark{d}} & \colhead{$B-R$} & \colhead{$B-V$} & \colhead{Sources}\\
\colhead{} & \colhead{[AU]} & \colhead{} & \colhead{[deg]} & \colhead{[AU]} &
\colhead{[magnitudes]} & \colhead{[magnitudes]} & \colhead{}}
\startdata
1998 $\rm WX_{24}$  & 43.612 & 0.035 &  0.9 &  42.097 & $1.79 \pm 0.07$ & $1.09 \pm 0.05$ & 6 \nl
1994 $\rm ES_{2}$   & 45.883 & 0.111 &  1.1 &  40.779 & $1.65 \pm 0.21$ & $0.71 \pm 0.15$ & 1 \nl
1994 $\rm VK_{8}$   & 43.133 & 0.031 &  1.5 &  41.800 & $1.68 \pm 0.07$ & $1.01 \pm 0.06$ & 6 \nl
1997 $\rm CU_{29}$  & 43.771 & 0.028 &  1.5 &  42.542 & $2.19 \pm 0.48$ & $1.22 \pm 0.10$ & 5, 7 \nl
1994 $\rm EV_{3}$   & 42.792 & 0.045 &  1.7 &  40.871 & $2.04 \pm 0.20$ & $1.50 \pm 0.19$ & 1 \nl
1995 $\rm WY_{2}$   & 46.873 & 0.128 &  1.7 &  40.857 & $1.65 \pm 0.21$ & $1.01 \pm 0.16$ & 1, 7 \nl
1992 $\rm QB_{1}$   & 44.092 & 0.072 &  2.2 &  40.907 & $1.59 \pm 0.14$ & $0.85 \pm 0.10$ & 1, 6, 7 \nl
1997 $\rm CS_{29}$  & 44.167 & 0.015 &  2.2 &  43.511 & $1.90 \pm 0.17$ & $1.10 \pm 0.04$ & 3, 5, 7 \nl
1997 $\rm CQ_{29}$  & 45.327 & 0.123 &  2.9 &  39.762 & $1.96 \pm 0.33$ & $0.99 \pm 0.09$ & 5, 7 \nl
1996 $\rm TK_{66}$  & 42.848 & 0.012 &  3.3 &  42.313 & $1.70 \pm 0.08$ & $1.04 \pm 0.07$ & 6, 7 \nl
1996 $\rm TS_{66}$  & 44.240 & 0.134 &  7.3 &  38.317 & $1.64 \pm 0.14$ & $1.02 \pm 0.05$ & 2, 3, 7 \nl
1999 $\rm DE_{9}$   & 55.891 & 0.423 &  7.6 &  32.242 & $1.51 \pm 0.04$ & $0.94 \pm 0.03$ & 7 \nl
1993 $\rm FW$       & 43.650 & 0.045 &  7.8 &  41.665 & $1.59 \pm 0.08$ & $0.95 \pm 0.06$ & 1, 5 \nl
1997 $\rm SZ_{10}$  & 48.190 & 0.366 & 11.8 &  30.546 & $1.79 \pm 0.09$ & $1.14 \pm 0.08$ & 6 \nl
1998 $\rm WH_{24}$  & 46.131 & 0.110 & 12.0 &  41.052 & $1.88 \pm 0.31$ & $0.94 \pm 0.04$ & 5, 6 \nl
1997 $\rm QH_{4}$   & 42.831 & 0.084 & 13.2 &  39.247 & $1.69 \pm 0.12$ & $1.03 \pm 0.10$ & 6, 7 \nl
1998 $\rm SM_{165}$ & 47.816 & 0.371 & 13.5 &  30.082 & $1.76 \pm 0.12$ & $1.01 \pm 0.10$ & 6 \nl
1997 $\rm CR_{29}$  & 47.310 & 0.216 & 19.1 &  37.082 & $1.36 \pm 0.27$ & $0.67 \pm 0.20$ & 7 \nl
\tablebreak
1996 $\rm TL_{66}$  & 84.917 & 0.587 & 23.9 &  35.033 & $1.02 \pm 0.11$ & $0.70 \pm 0.04$ & 2, 3, 4, 7 \nl
1999 $\rm OY_{3}$   & 43.631 & 0.166 & 24.3 &  36.407 & $1.08 \pm 0.02$ & $0.71 \pm 0.01$ & 6 \nl
1999 $\rm KR_{16}$  & 48.477 & 0.298 & 24.9 &  34.006 & $1.84 \pm 0.06$ & $1.10 \pm 0.05$ & 7 \nl
1996 $\rm TO_{66}$  & 43.364 & 0.115 & 27.4 &  38.372 & $1.07 \pm 0.05$ & $0.69 \pm 0.03$ & 2, 3, 4, 7 \nl
1996 $\rm RQ_{20}$  & 44.036 & 0.104 & 31.6 &  39.467 & $1.54 \pm 0.17$ & $0.96 \pm 0.13$ & 7 \nl
\enddata
\tablecomments{Objects are presented in order of increasing
inclination.  Due to space considerations, Plutinos have been excluded
from this table.  Orbital elements were provided by the Minor Planet
Center.}
\tablenotetext{a}{semimajor axis}
\tablenotetext{b}{eccentricity}
\tablenotetext{c}{$i$ refers to the inclination with respect to the
ecliptic.  Transforming the inclination to the invariable plane
(mean correction 0.6\arcdeg, Burkhardt 1982) does not have a
significant impact on results.}
\tablenotetext{d}{perihelion distance}
\tablenotetext{1}{Luu \& Jewitt 1996}
\tablenotetext{2}{Jewitt \& Luu 1998}
\tablenotetext{3}{Tegler \& Romanishin 1998}
\tablenotetext{4}{Barucci et al. 1999}
\tablenotetext{5}{Barucci et al. 2000}
\tablenotetext{6}{Tegler \& Romanishin 2000}
\tablenotetext{7}{Jewitt \& Luu 2001}
\label{otherdata}
\end{deluxetable}
\end{center}

\clearpage

%\section*{Figure Captions}

\clearpage

\clearpage

\end{document}